\documentclass{aa} 
\usepackage{graphicx}
\usepackage{natbib}
\bibpunct{(}{)}{;}{a}{}{,}
\usepackage{txfonts}
\usepackage{array}
\usepackage{graphics}
\usepackage{latexsym}
\usepackage{amssymb}
\usepackage{amsmath}
\usepackage{fancyhdr}
\usepackage{float}
\usepackage{morefloats}
\usepackage{slashbox}
\usepackage{multirow}
\usepackage[toc,page]{appendix}
\bibpunct{(}{)}{;}{a}{}{,}
\include{hyphe}

\usepackage[english]{babel}

\begin{document}
\title{T-PHOT version 2.0: improved algorithms for background subtraction, local convolution, kernel registration, and new options}

\author{E. ~Merlin \inst{1} 
\and N. ~Bourne \inst{2}
\and M. ~Castellano \inst{1}
\and H.~ C. ~Ferguson \inst{3}
\and T. ~Wang \inst{4}
\and S. ~Derriere \inst{5}
\and J. ~S. ~Dunlop \inst{2}
\and D. ~Elbaz \inst{4}
\and A. ~Fontana \inst{1}
}
\institute{INAF - Osservatorio Astronomico di Roma, Via Frascati 33, I - 00078 Monte Porzio Catone (RM), Italy
\email{emiliano.merlin\char64oa-roma.inaf.it}\label{inst1}
\and SUPA\thanks{Scottish Universities Physics Alliance}, Institute for Astronomy, University of Edinburgh, Royal Observatory, Edinburgh, EH9 3HJ, U.K. \label{inst2}
\and Space Telescope Science Institute, 3700 San Martin Drive, Baltimore, MD 21218, USA \label{inst3}
\and Laboratoire AIM-Paris-Saclay, CEA/DSM/Irfu - CNRS - Universit\'e Paris Diderot, CEA-Saclay, pt courrier 131, F-91191 Gif-sur-Yvette, France \label{inst4}
\and Observatoire astronomique de Strasbourg, Universit\'e de Strasbourg, CNRS, UMR 7550, 11 rue de l'Universit\'e, F-67000 Strasbourg, France \label{inst5} 
}

\abstract
{}
{
We present the new release v2.0 of \textsc{t-phot}, a publicly available software package developed to perform PSF-matched, prior-based, multiwavelength deconfusion photometry of extragalactic fields.
}
{
New features included in the code are presented and discussed: background estimation, fitting using position dependent kernels, flux prioring, diagnostical statistics on the residual image, exclusion of selected sources from the model and residual images, individual registration of fitted objects.
}
{
The new options improve on the performance of the code, allowing for more accurate results and providing useful aids for diagnostics.
}
{}
\authorrunning{E. Merlin et al.}
\titlerunning{T-PHOT v2.0}
\maketitle

\section{Introduction}

\textsc{t-phot} \citep[][M15 hereatfer]{Merlin2015} is a public software package designed to perform precision photometry on a low resolution extragalactic image (LRI) using the information given by priors obtained from a higher resolution image (HRI) of the same field. It has been developed and released within the \textsc{Astrodeep} project and it is being increasingly used in the community.

This paper presents the features included in the new publicly released version 2.0.  The new package is downloadable from the webpage \textit{http://www.astrodeep.eu/t-phot}. Version 2.0 is back-compatible with the last publicly released version, 1.5.11: the installation procedure and the required input have not changed, and old parameter files can be used. All the features present in v1.5.11 are still available.

A detailed description of \textsc{t-phot} and its capabilities and limitations is given in M15; for the reader's convenience, we give a brief review of the code philosophy here. Starting from spatial and morphological information on a list of priors, \textsc{t-phot} produces low-resolution models (\textit{templates}) by means of a convolution kernel, and assignes to each normalized model a multiplicative factor, to match the global observed flux in the LRI. This technique is particularly useful to disentagle the contribution to the observed flux coming from blended sources. 

The search for the fluxes in the LRI is performed solving a linear system
\begin{equation}
I = F_1P_1+...+F_NP_N 
\end{equation}

\noindent where $I$ contains the pixel values of the fluxes in the LRI, $P_i$ are the normalized fluxes of the templates for the $N$ objects in the (region of the) LRI being fitted, and $F_i$ are the multiplicative scaling factors for each object. In physical terms, $F_i$ represent the flux of each object in the LRI (that is, it is the unknown to be determined).

In the Gaussian additive noise regime (a condition typically satisfied in astrophysical infrared images), the best fit for the unknown fluxes can be derived by minimizing a $\chi^2$ statistic, 
\begin{equation}
\chi^2=\sum_{m,n}\left[ \frac{I(m,n)-M(m,n)}{\sigma(m,n)} \right]^2
\end{equation}

\noindent where $m$ and $n$ are the pixel indexes,
\begin{equation}
M(m,n)=\sum_i^N F_i(m,n)P_i(m,n)
\end{equation}

\noindent and $\sigma$ is the RMS value in the pixel.

In practice, the problem is reshaped into a matrix equation 
\begin{equation}
AF=B \label{linsys}
\end{equation}
\noindent where the matrix $A$ contains the coefficients $P_i P_j / \sigma^2$, $F$ is the vector of the unknowns, and $B$ is a vector given by $I_i P_i / \sigma^2$ terms.

The system can be solved at once on the whole image, or on portions of the LRI, either determined by an arbitrary regular grid of cells of with cells centered on each object.

\textsc{t-phot} can process simultaneously three types of priors: real cutouts from a high-resolution image, analytical 2-D models, or point-sources. The pipeline of \textsc{t-phot} consists of ``stages'', each of which performs a well defined task; the best results are usually obtained performing two runs (\textit{pass1} and \textit{pass2}), the second one using local convolution kernels, registered after the $X,Y$ local shifts are determined during \textit{pass1}. 

The pipeline can be specified using the keyword \texttt{order} in the parameter file. The simplest way to run the code is to simply set \texttt{order standard} for a typical \textit{pass1} run, and \texttt{order standard2} for the subsequent \textit{pass2} run.

In v2.0 it is also possible to set \texttt{order FIRstandard} and/or \texttt{order FIRstandard2} for typical far-infrared (FIR) processing, i.e. using only point-like priors and PSF-shaped templates for the fit. If this option is used, any input given for high resolution real priors or model priors will be ignored.

\textsc{t-phot} outputs a catalogue with IDs, positions, measured fluxes and corresponding uncertainties for each source in the priors list, as well as a number of useful diagnostics (flags, covariance indexes, residual maps, etc.).

\begin{figure}[t!] 
\includegraphics[width=8cm]{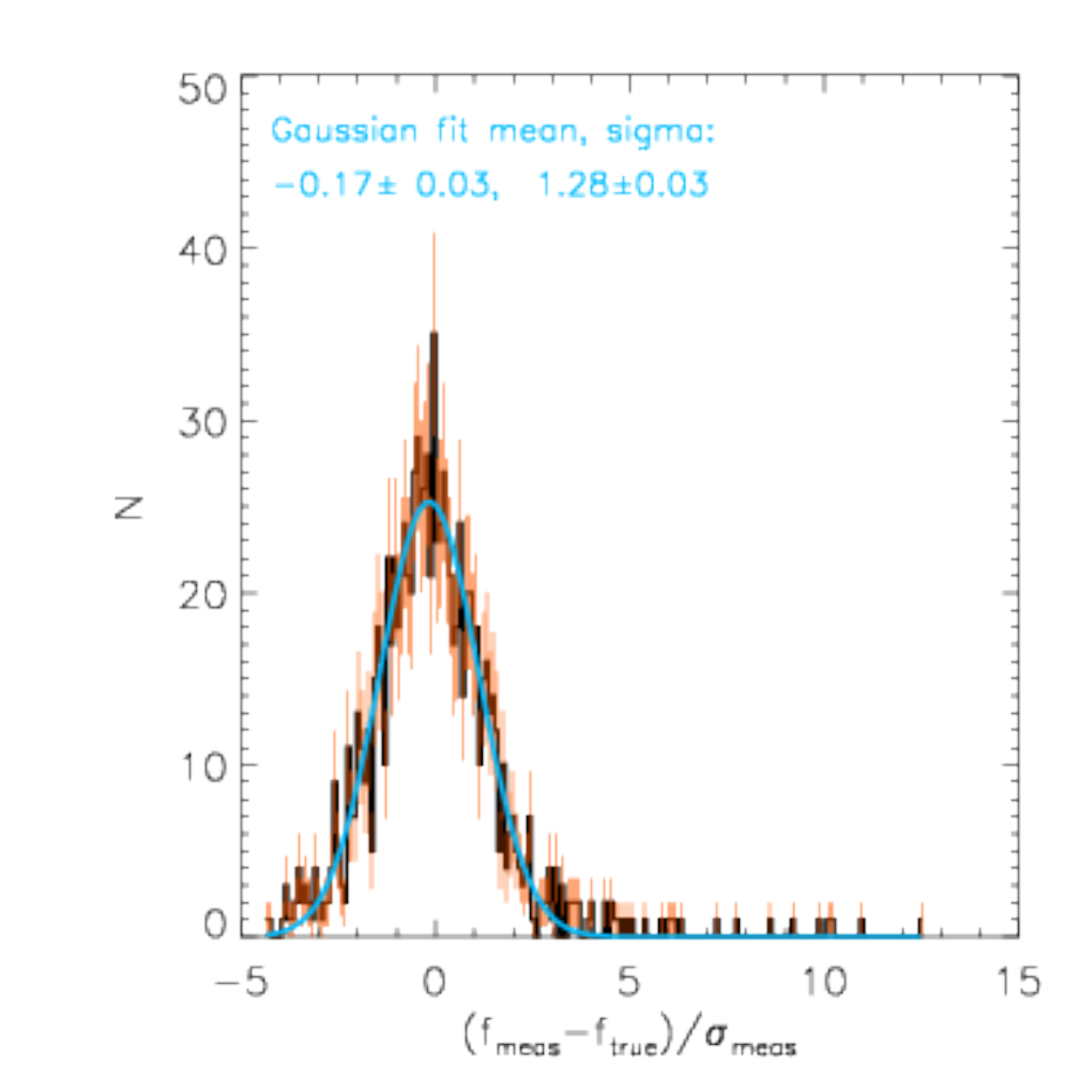}
\centering
\caption{Global background subtraction. 1000 fake sources have been injected on a SCUBA-2 450 $\mu$m map, previously background subtracted with the \textsc{t-phot} routine, and the output measured flux for each source has been compared to the input ``true'' flux. The average is consistent with zero, within the uncertainties of the method. See Bourne et al., 2016 in preparation.}\label{globbkg}
\end{figure}

\section{New options available in T-PHOT v2.0}

The new features available in v2.0 are the following.

\begin{itemize}
  \item \textbf{Background estimation}, with two methods: global subtraction of a constant fitted value, and local fit of individual “background templates”. See Section \ref{background}.
  \item \textbf{Local/individual kernel fitting}: it is possible to associate a different kernel to each source to optimize the fit, coping with local variations of the PSFs. See Section \ref{indker}.
  \item \textbf{Individual source registration (\textit{dance})}: after the fit, a refinement of the spatial registration of the objects is performed on individual basis rather than on arbitrary regions. See Section \ref{inddance}.
  \item \textbf{Flux prioring}: the flux of selected or all sources can be constrained to a given desired value, within a chosen uncertainty limit. See Section \ref{fluxprior}.
  \item \textbf{Statistics on the residuals}: the output includes a new text file with diagnostic statistics for each source, based on the residual image produced after the fit. See Section \ref{residstats}.
  \item \textbf{RMS uncertainty threshold to exclude sources from the fit}: if a source includes a pixel with RMS uncertainty exceeding a chosen value, it will be excluded from the fitting procedure. See Section \ref{rmscheck}.
  \item \textbf{Model building with selected sources}: it is possible to build a model image (and a residual image) including only a selection of sources from the priors list. See Section \ref{diags}.
\end{itemize}

A further, technical new option is the command line input: it is now possible to enter parameters from command line, in case over-writing the ones specified in the parameter file. Keywords and corresponding values can be entered typing \texttt{-<keyword> <value>} after the parameter file specification.

A revision of the code architecture in the sources registration and in the convolution stages has also been performed, to make the workflow simpler and better organized. The low-resolution templates are now registered on the fly during the second pass convolution stage, and the second pass local kernels are not stored anymore on the local hard disk.

The new options are described in details in the documentation included in the software tarball. In the following subsections we present them in summary.

\begin{figure*}[t!] 
\includegraphics[width=6cm]{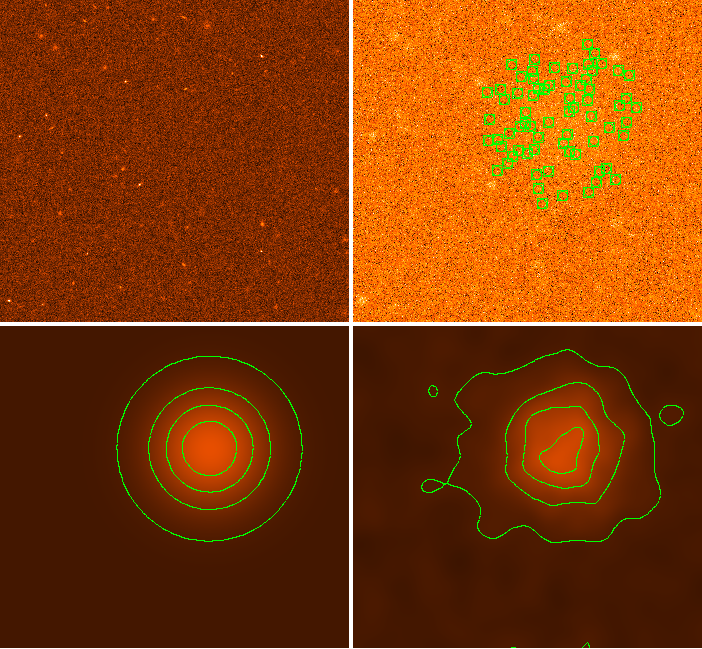}
\includegraphics[width=9cm]{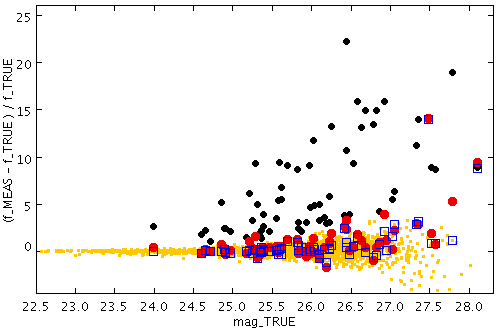}
\centering
\caption{Left panels: simulated images to test the local background estimation techinque. Top left: HRI (FWHM=0.2''); top right: LRI (FWHM=1.66''; many sources are too faint to be seen); bottom left: artificially added background (on a different scale and with density countours to enhance visibility); bottom right: background estimated by \textsc{t-phot}, after smoothing with a gaussian kernel. The plot on the right shows the relative error on the measured flux versus the input true magnitude: yellow tiny dots refer to the whole catalogue in a run with no background enhancement; in the run with the enhanced background but without applying the local background estimation method, many sources (black dots) have largely overestimated fluxes (their positions are shown with green marks in the LRI image, second subpanel of the left image): they are all gathered where the artificial background is stronger). When applying the local background estimation method, the measured fluxes of these sources are much more reasonable (red dots), and close to their values measured in the run without background enahncement (empty blue squares).}\label{bksub_sim}
\end{figure*}

\subsection{Background estimation} \label{background}

\textsc{t-phot} v2.0 can estimate a constant background from the whole image, and/or a local background for each source, during the fitting stage. 

\subsubsection{Global background}

To estimate a constant global background, the keyword \texttt{fitbackground} must be set to \texttt{true} in the parameter file. An additional constant term will be added to the linear system matrix. It is important to note that:
 
\begin{itemize}
\item it is strongly recommended to use this option only when fitting the whole image at once. If a cell method is used for the fit, the local background will be computed for each cell and this might lead to unphysical patchy solutions;
\item the value of the background will only be output in the log file of the fitting routine, while the model and residual images will \emph{not} be background subtracted; to visualize the results, the user must subtract the fitted value from the residual image. On the contrary, the fitted fluxes in the output catalog obviously take into account the background and must not be corrected.
\end{itemize}

The global background subtraction has been tested extensively in \textit{Herschel} and SCUBA2 images in single-fit mode. Fig. \ref{globbkg} shows the results of one such test in which a SCUBA-2 450 $\mu$m image of the COSMOS-CANDELS field was fitted with a prior catalogue from \textit{K} band and IRAC 3.6 $\mu$m priors (Bourne et al. 2016, in preparation). To test the background subtraction, a single fake source was added to the image and prior catalogue at a random position, with its flux drawn from a uniform logarithmic distribution between 0.01-10 mJy. \textsc{t-phot} was used to extract the fluxes of all priors including the artificial source, and the procedure was repeated 1000 times. The distribution of the difference between measured and true fluxes of the artificial sources in the 450 $\mu$m image is shown in Fig. \ref{globbkg}. By fitting a gaussian profile to this histogram, we find that the mean offset is a small fraction of the output uncertainties, while the width of the distribution is marginally larger than the measurement uncertainties by a factor 1.3. We found no significant trend in this flux offset as a function of input flux. The mean of the unbinned data ($f_\text{meas}-f_\text{true}$) is $0.15\pm0.18$ mJy, while the variance-weighted mean is $-0.02\pm0.05$ mJy; both are consistent with zero, indicating that background subtraction is successful and fluxes over a wide range can be recovered reliably without bias.

\begin{figure*}[t!] 
\includegraphics[width=18cm]{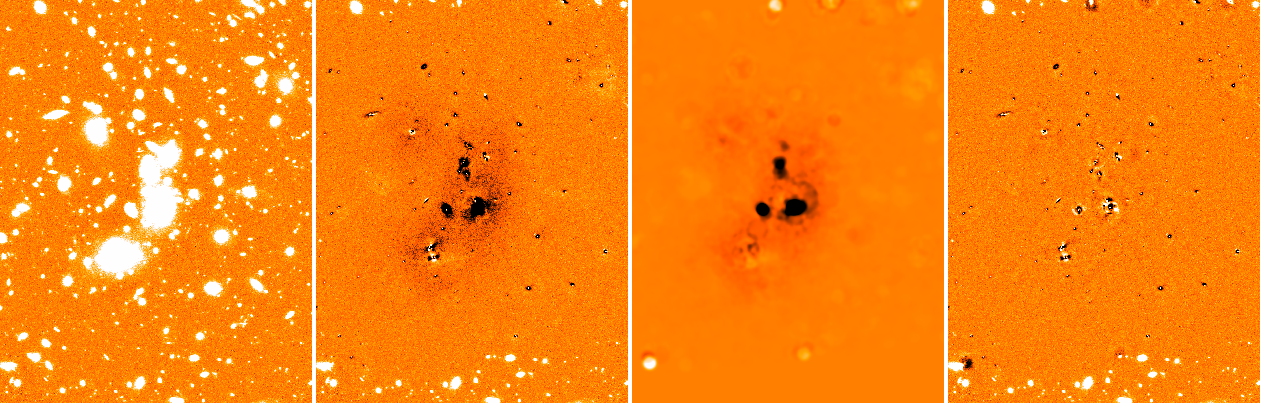}
\centering
\caption{Local background subtraction. Left to right: original LRI image (Abell 2744 $Ks$ band, courtesy G. Brammer); residual after standard fitting; background model, build as a smoothed collage of the fitted local background templates; residual after fitting on background subtracted image.}\label{bksub}
\end{figure*}

\subsubsection{Local background}

If the background is expected to vary strongly within the fitted region, it might be useful to fit a locally varying background. To do so, \textsc{t-phot} can create an additional, flat fake object for each real source, and add it to the priors list; these fake ``background'' objects are then fit simultaneously with the real objects, providing an estimation of a possible flat background flux. To switch on this option, the keyword \texttt{fit\_loc\_bkgd} in the paramfile must be set to an integer, giving the offset to be assigned to the IDs of the fake bakground templates (it should be larger than the maximum ID of real objects). 

Important notes:
\begin{itemize}
\item the fitted value for these background templates may scatter significantly from a reliable average value. Therefore, only an average of all the ``background'' templates will give a reasonable estimate of the background.
\item including these templates will change the covariance matrix and hence affect the error budget of the measured fluxes.
\end{itemize} 
To cope with these issues, it is strongly recommended to build a model image including only the fitted background templates (see Sect. \ref{diags}), subtract it from the real LRI (possibly after some smoothing), and repeat the fit in standard mode on this background subtracted LRI.

Fig. \ref{bksub_sim} shows the results of a test on a simulated set of images. An artificial, gaussian-shaped background light has been added to the original LRI; fluxes have been measured with a standard run, and then with a run including the local background subtraction algorithm. While in the first case the fluxes of the sources in the region where the background has been enhanced turn out to be largely wrong, the new method allows for a good recovery of their true flux.

Fig. \ref{bksub} shows the effect of the application of this technique for the fit of the Abell 2744 cluster in the $Ks$ band, as described in \citet{Merlin2016}.

\begin{figure*}[t!] 
\includegraphics[width=18cm]{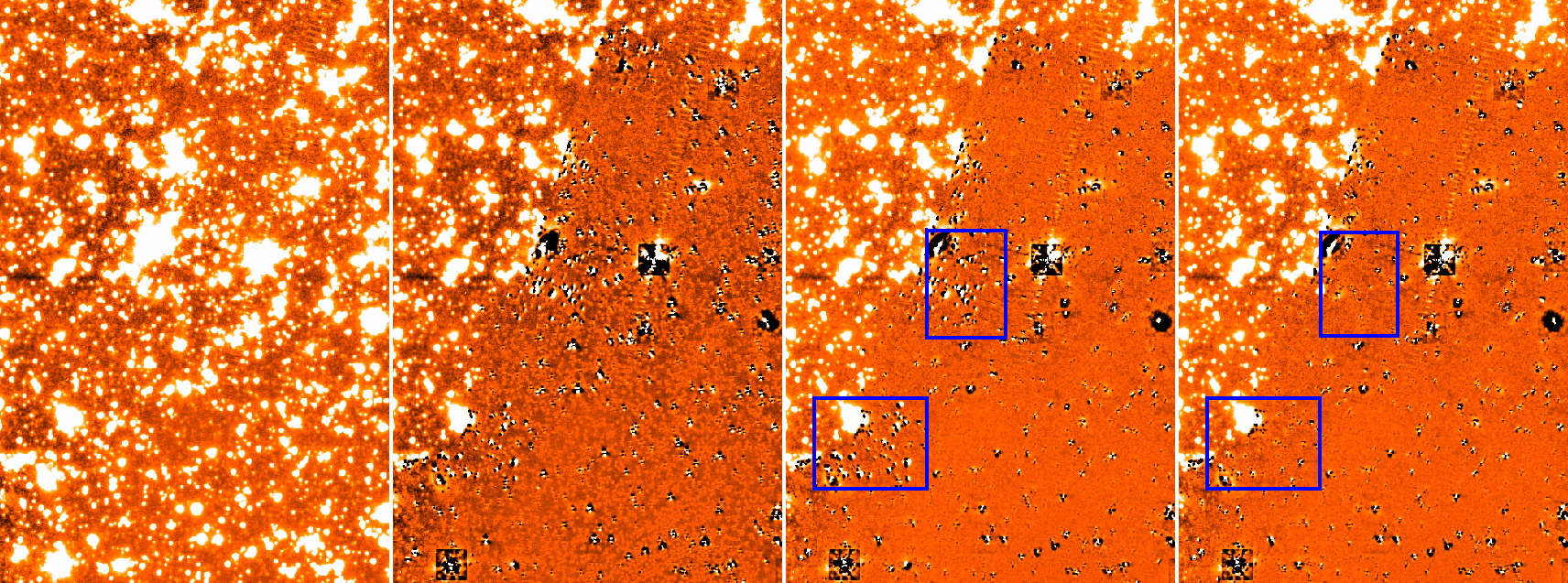}
\centering
\caption{From left to right: original LRI (a portion of GOODS-South observed with IRAC at 3.6 $\mu$m, courtesy R. McLure); residual after standard fitting using a single convolution kernel; residual after fitting using a different individual kernel for each source (Sect. \ref{indker}), tailored on the basis of the positional angles of the pointings used to build the mosaic (global background subtraction has also been applied, see Sect. \ref{background}); residual after including the individual kernel registration (Sect. \ref{inddance}; in the last two panels, blue boxes highlight regions in which the the improvement using this technique is evident).}\label{scandels}
\end{figure*}

\begin{figure*}[t!] 
\includegraphics[width=14cm]{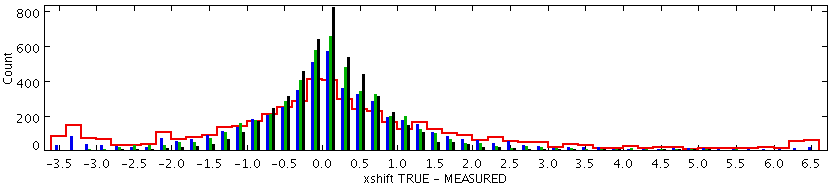}
\includegraphics[width=14cm]{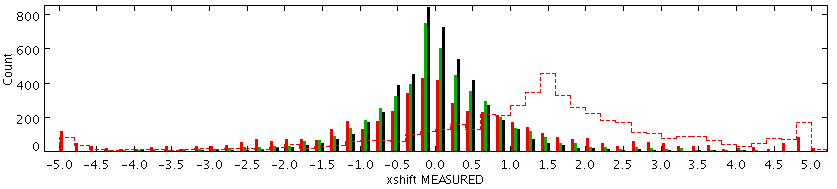}
\centering
\caption{Individual registration of sources (dance). Top panel: histogram of the difference between true and measured shifts in the $X$ direction in a simulated image; red: region-based dance (v1.5.11); blue bars: individual dance, without smoothing; green bars: individual dance with smoothing on a fixed $R_{neigh}$ area; black bars: individual dance with smoothing on a fixed number $N_{neigh}=100$ of neighbouring sources. Bottom panel: histogram of the measured shifts in the $X$ direction in a simulated image where no shifts had been applied; red bars: region-based dance (v1.5.11); green bars: individual dance with smoothing on a fixed $R_{neigh}$ area; black bars: individual dance with smoothing on a fixed number $N_{neigh}=100$ of neighbouring sources; dashed red: region-based dance on the image \emph{with} shifts, for reference. See text for more details.}\label{dance_hist}
\end{figure*}

\subsection{Local or individual kernels fitting} \label{indker}

PSFs can vary substantially in different regions of the same image. Provided this variation can be characterized quantitatively, it is possible to input a list of individual convolution kernels or PSFs, each one to be associated to one prior (of course, it is possible to link the same kernel to more than one prior, e.g. characterizing regions rather than individual sources). \textsc{t-phot} will process each object using the corresponding convolution kernel.

An example of the effectiveness of this approach is depicted in Fig. \ref{scandels}. The first three panels show, from left to right: the LRI (in this case, a portion of the GOODS-South field in the IRAC 3.6 $\mu$m band); the residuals obtained using a single PSF on the whole image; and the residuals obtained using individual kernels for each source (created averaging model PSFs, each one rotated accordingly to the position angle of the observations and weighted by the exposure time of each pointing\footnote{We neglected shear and rescaling of the PSFs in this case.}), plus the global background subtraction technique described in Sect. \ref{background}. It is clear that the individual kernel fitting yields much more accurate results. 

\subsection{Individual sources registration} \label{inddance}

After the fitting stage, a spatial registration procedure (the \texttt{dance} stage) can be performed to mitigate the effects of possible local astrometric imprecisions. The procedure is based on a cross-correlation between the models collage and the original LRI. In v1.5.11, this cross-correlation is made on the basis of a predifined subdivision of the LRI in a regular grid of cells; on the contrary, in v2.0 the process is performed on a source by source basis. The cross-correlation is made on the pixels belonging to the area of the low-resolution template of each source (or on a minimum user-defined predefined area if the template is too small). To avoid local numerical divergencies, the final values of the $X,Y$ shifts are smoothed computing a weighted mean including $\simeq N_{neigh}$ nearest neighboring sources, or all the neighbours within a given $R_{neigh}$. The weight of each neighbour is computed as the ratio between the peak value of its own correlation function (the highest this value, the more reliable the evaluation of the shift for this source) and the distance from the central object.

\begin{figure}[t!] 
\includegraphics[width=8cm]{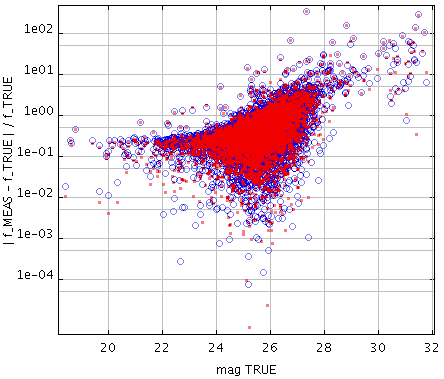}
\centering
\caption{Individual registration of sources (dance). Absolute difference between true and measured fluxes in the same simulated image of Fig. \ref{dance_hist}, where no shifts are present, before (red dots) and after (blue open circles) applying the individual smoothed registration procedure. See text for more details.}\label{dance_fluxes}
\end{figure}

While slightly more time consuming, this method allows for a much more precise determination of the spatial shifts needed to register each model, as it is evident looking at the third and fourth panels of Fig. \ref{scandels} (blue boxes highlight the regions in which the improvement is more evident). 
We performed a test to make sure that this method increases the accuracy while not introducing biases. We produced two simulated LRIs from the same input catalogue: the first one with each source shifted by a known amount in the $X$ and $Y$ directions, and the second one without shifts. 
In the upper panel of Fig. \ref{dance_hist} we plot the histograms of the difference between the true and the measured shifts in the $X$ direction on the first LRI, with four methods: the region-based dance (used in v1.5.11, red histogram), and the individual dance without (blue bars) and with smoothing (green bars: smoothing on all neighbors within $R_{neigh}$; black bars: smoothing on the closest $N_{neigh}=100$ neighbors). The smoothed individual dance with fixed $N_{neigh}$ yields the best results, reducing spurious large shifts and increasing the number of sources with a good estimation of the true shift. Similar results are obtained for the $Y$ shifts. 
In the bottom panel of Fig. \ref{dance_hist} we show the measured shifts in the $X$ direction on the second LRI, where no true shift is present: again, the smoothed individual dance with fixed $N_{neigh}$ (green and black bars) yields more accurate results than the region-based dance (red bars; the red dashed histogram is, for reference, the measured shift in the first LRI). 
Similar results are obtained for the $Y$ shifts.
Finally, to make sure that the new method does not introduce biases in the photometry, we checked that the measured fluxes are consistent with their true values. We did so both on the second LRI of the previous test (see Fig. \ref{dance_fluxes}), and on a new image with only PSF-shaped objects displaced on a regular grid to avoid contaminations and mismatches. We find that while a straightforward individual registration can, in some cases, slightly affect the accuracy of the photometry, because the noise can lead the local cross-correlation process (depending on the extension of the objects under consideration), virtually no bias is introduced when the smoothing approach is adopted.

From all these tests, it turns out that the optimal registration technique depends on the particular case under analysis. In the considered idealized simulations, where the artificial shifts have been added analytically and follow a smooth pattern, the accuracy in determining the true shifts keeps increasing as $N_{neigh}$ increases, although if no shifts are present an even better result is obtained smoothing on a fixed area rather than keeping the number of neighbours constant. However, in real cases the pattern of shifts is generally chaotic, likely with abrupt discontinuities over the observed fields. This makes it difficult to foresee a general good-practice standard. We therefore choose to leave $N_{neigh}$ as a free input parameter, also including the possibility to smooth over fixed $R_{neigh}$.


\subsection{Flux prioring} \label{fluxprior}

In v2.0 it is possible to perform the fitting routine enabling an option to constrain measured fluxes around a chosen fixed value, with a given allowed uncertainty. This can be useful to constrain sources on the basis of a SED-fitted predicted flux in the measurement passband.

To do so, the matrix $A$ and the vector $B$ of the linear system in Eq. \ref{linsys} are modified as follows:
\begin{itemize}
\item $B_i$ becomes $B_i + f_i/\sigma_i^2$, and
\item when $i=j$, the element $A_{ij}$ changes to $A_{ij} + 1/\sigma_i^2$, 
\end{itemize}
\noindent where $f_i$ is the estimated flux for source $i$ that has
to be used as prior for that source, and $\sigma_i$ is its associated 
uncertainty (this procedure is known as L2 regularization).

To enable this ``flux prioring'' option, a text file must be prepared, in which each source is associated with its prior flux and the relative allowed uncertainty, along with a flag indicating whether the prior must be used or not.

\begin{figure}[t!] 
\includegraphics[width=8.5cm]{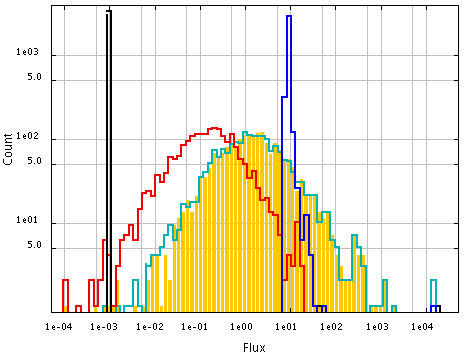}
\centering
\caption{Results of a series of tests on a simulated image using the flux prioring method. Different histograms refer to separate runs on the same LRI. The yellow bars are the fluxes $f$ measured without imposing any constraints (run $A$). In all the other runs, a given set of flux priors with a corresponding uncertainty has been given to all the sources, and the histograms show the measured fluxes: $f_{constr,B}=f \pm 0.5$ (cyan), $f_{constr,C}=10 \pm 0.1$ (blue), $f_{constr,D}=10^{-3} \pm 0.1$ (red), $f_{constr,E}=10^{-3} \pm 10^{-4}$ (black). Note that: (i) the run without constraints (yellow) yields very similar results to the run where each source has its ``true'' input flux as prior (B, cyan); (ii) in the other runs the flux is forced to a fixed value which is retrieved consistently with the allowed uncertainties. See text for more details.}\label{fluxpr}
\end{figure}

Fig. \ref{fluxpr} shows the results of a test on a simulated image, on which five different runs were performed: 
\begin{itemize}
\item (i) a reference run $A$, without constraints on the fluxes (yellow histogram), 
\item (ii) a run $B$ in which the fluxes of all sources were constrained to be consistent with their input ``true'' fluxes within $\sigma_{constr}=0.5$ (cyan), 
\item and three other runs in which the measured fluxes were forced to arbitrary values with different error budgets: 
\begin{itemize}
\item (iii) $f_{constr}=10$ and $\sigma_{constr}=0.1$ (blue, run $C$), 
\item (iv) $f_{constr}=10^{-3}$ and $\sigma_{constr}=0.1$ (red, run $D$), and 
\item (v) $f_{constr}=10^{-3}$ and $\sigma_{constr}=10^{-5}$ (black, run $E$). 
\end{itemize}
\end{itemize}

The results show that the output fluxes are always consistent with the expectations: in particular, the fluxes from run $A$ and $B$ are very similar; and runs $C$ to $D$ all yield fluxes consistent with the required prior flux, the amplitude of the scatter depending on the allowed uncertainty (in particular, run $D$ yields fluxes closer to the ``true'' values than to the given prior flux, because of the large allowed $\sigma_{constr}$, while in run $E$ all sources have $f_{meas} \simeq f_{constr}$ because $\sigma_{constr}$ is small)\footnote{The input $\sigma$ for each source gives an estimate of the reliability of the corresponding prior, as an additional term in the system matrix, but the measured flux is not unavoidably forced to stay within its limit. This is the reason why many sources end up having measured fluxes with a scatter larger than $3 \sigma$.}. In realistic cases, one might want to constraint the flux of a subset of sources to some predicted value (e.g. using SED-fitting techniques); this is similar to our case $B$.

\begin{figure}[t!] 
\centering
\includegraphics[width=5.5cm]{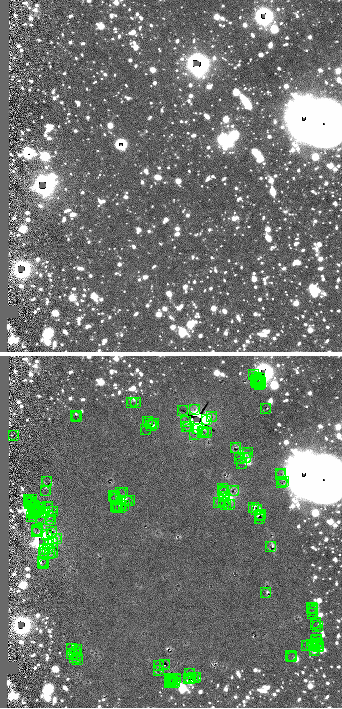}
\includegraphics[width=7cm]{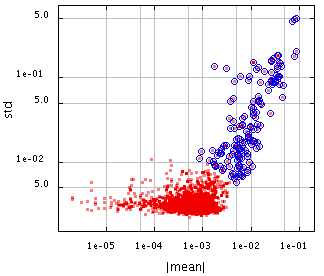}
\centering
\caption{Original LRI (upper) and residual after fitting (center) on a FORS2 $R$-band image; the sources with green squares correspond to the ones selected by eye as deviating from the global distribution in the bottom plot (the blue open circles; red dots are the whole catalogue), where the standard deviation of the residual of each source has been plotted against the absolute value of the mean of the residual (both values have been taken from the output file from \textsc{t-phot}).}\label{residstat}
\end{figure}

\subsection{Statistics on the residual image} \label{residstats}

A file containing diagnostic statistics for each source, computed on the residual image, can be output from this version. A file will be produced listing mean, median, RMS and kurtosis computed on the pixels of the residual image belonging to the template model. Also, the same values computed only on an inner and outer regions (the limit of such regions is defined as the radius at which the flux of the template is half the value of the peak) will be output.

Fig. \ref{residstat} shows how this feature can be a useful aid to analyse the reliability of the results, e.g. to single out sources with high standard deviation in the pixel fluxes on the residual image.

\subsection{Exclusion of high RMS uncertainty sources from the fit} \label{rmscheck}

In v2.0 it is possible to include a keyword in the parameter file to exclude from the fit sources belonging to regions with exceedingly large RMS uncertainty values (e.g. flawed regions, or artificially enlarged borders). If the value of the keyword \texttt{rmscheck} is set equal to some $c_{RMS}>0$, a check is performed on the RMS map and sources having their central pixels with a value higher than $c_{RMS}$ are automatically excluded from the list of the sources to be fitted. These sources will be re-included in the final output catalog, with ``99.0'' and zero values in the relevant fields.

\subsection{Model and residual images production}\label{diags}

After the fitting procedure, \textsc{t-phot} produces a final catalogue with the determined fluxes, and two diagnostic images: a model image obtained producing a collage with the low resolution templates of the sources, each one put at its correct position and multiplied by its fitted flux; and a residual image, obtained subtracting the model image from the original LRI.

In v2.0 it is possible to feed the code with a file containing a list of IDs from the HRI catalog to be \emph{excluded} from the model image (they will therefore remain unsubtracted in the residual image). This feature can be useful to isolate objects removing neighbors, or to remove bright foreground sources leaving background objects.

\section{Conclusions}

We have presented and discussed the new options implemented in \textsc{t-phot} v2.0: background estimation, fitting using individual kernels,  individual registration of fitted objects, flux prioring, statistics on the residual image, exclusion of selected sources from the model and residual images. 

The code is publicly available for downloading from the \textsc{Astrodeep} website.

\bibliographystyle{aa}
\bibliography{mnemonic,biblio}

\end{document}